\documentclass[11pt,a4paper]{article}
\pdfoutput=1
\usepackage{jheppub}
\usepackage{rotating}
\usepackage{wrapfig,enumitem}
\usepackage{amsfonts}
\usepackage{amsmath}
\usepackage{slashed}
\usepackage{amssymb}
\usepackage{latexsym}
\usepackage{multirow}
\usepackage{hyperref}
\hypersetup{colorlinks=true,citecolor=red,linkcolor=magenta,urlcolor=blue}
\usepackage{relsize}
\usepackage{booktabs}
\usepackage{subfigure}
\usepackage[capitalise]{cleveref}
\usepackage{fancyhdr}
\usepackage{float}
\usepackage{graphicx}
\usepackage{microtype}
\usepackage{tabularx}
\usepackage{xcolor}
\usepackage{adjustbox,bm}
\usepackage[bottom]{footmisc}
\usepackage{orcidlink}
\usepackage{physics}

\newcommand{\mh}{m_h}
\newcommand{\mH}{m_H}
\newcommand{\mA}{m_A}
\newcommand{\mHp}{m_{H^\pm}}
\newcommand{\mbar}{\bar{m}}
\newcommand{\tb}{\tan\beta}
\newcommand{\cbma}{c_{\beta-\alpha}}
\newcommand{\sbma}{s_{\beta-\alpha}}
\newcommand{\kala}{\kappa_\lambda}

\newcommand{\laphi}{\lambda_{h\phi\phi}}

\newcommand{\lahHH}{\lambda_{hHH}}
\newcommand{\lahAA}{\lambda_{hAA}}
\newcommand{\lahHpHp}{\lambda_{hH^+H^-}}

\newenvironment{cedescription}{%
   
   \begin{description}[leftmargin=0.5cm, style=sameline]%
}{%
   \end{description}%
}
\newcolumntype{C}{>{$}c<{$}}
\newcolumntype{L}{>{$}l<{$}}
\newcolumntype{R}{>{$}r<{$}}

\title{Exploring Extended Higgs Dynamics via Higgsstrahlung at FCC-ee}
\preprint{KA-TP-18-2026, P3H-26-070, DESY-26-113}
\author[a,b]{Anisha\orcidlink{0000-0002-5294-3786},}
\author[c]{Francisco Arco\orcidlink{0000-0003-3651-1788},}  
\author[a]{Stefano Di Noi\orcidlink{0000-0002-1140-4073},}
\author[d]{Christoph Englert\orcidlink{0000-0003-2201-0667},} 
\author[a]{and Margarete M\"uhlleitner\orcidlink{0000-0003-3922-0281}}
\affiliation[a]{Institute for Theoretical Physics, Karlsruhe Institute of Technology,\\76131 Karlsruhe, Germany}
\affiliation[b]{Institute for Astroparticle Physics, Karlsruhe Institute of Technology,\\76344 Eggenstein-Leopoldshafen, Germany}
\affiliation[c]{Deutsches Elektronen-Synchrotron DESY, Notkestr. 85, 22607 Hamburg, Germany}
\affiliation[d]{Department of Physics \& Astronomy, University of Manchester, Oxford Road,\\Manchester M13 9PL, United Kingdom}
\emailAdd{anisha@kit.edu}
\emailAdd{francisco.arco@desy.de}
\emailAdd{stefano.dinoi@kit.edu}
\emailAdd{christoph.englert@manchester.ac.uk}
\emailAdd{margarete.muehlleitner@kit.edu}

\abstract{The Future Circular Electron-Positron Collider (FCC-ee) will probe aspects of the Higgs boson and the electroweak scale with unprecedented precision via associated Higgs production with a $Z$ boson. We focus on the Two-Higgs-Doublet Model (2HDM) to frame the FCC-ee's precision constraints within a small, concrete parameter space at next-to-leading order. We demonstrate that the projected precision of the FCC-ee's 240 GeV run will enable a precise analysis of Beyond-the-Standard-Model (BSM)-relevant Higgs-sector interactions near the alignment limit. As the key aspects of the 2HDM that drive deviations of the cross section from its Standard Model expectation (presence of new scalar states, Higgs mixing, and non-trivial inter-Higgs couplings) are also typically present in extensions more complex than the 2HDM, this demonstrates the FCC-ee’s indirect potential for unveiling BSM physics around the electroweak scale.}
\begin{document}
%%%%%%%%%%%%%%%%%%%
\maketitle
\flushbottom
\allowdisplaybreaks
%%%%%%%%%%%%%%%%%%%
\section{Introduction}
\label{sec:intro}
%%%%%%%%%%%%%%%%%%%
Experimental indications that the Standard Model (SM) is incomplete are ubiquitous. Meanwhile, concrete theories, whether ultraviolet-complete or not, are being put under increasing data-driven pressure, especially through explorations of particle dynamics at the highest experimentally attainable energies at the Large Hadron Collider (LHC). This raises the question of whether the solutions to the SM's shortcomings are indeed phenomenologically linked to experimentally accessible features of the weak scale. Currently, no direct evidence for new interactions beyond the SM exists in the high-energy particle physics domain. Several anomalies are currently observed; however, they do not seem to point to a more profound theory directly linked to the shortcomings of the SM. 

Since the discovery of the Higgs boson completed the SM particle spectrum, any attempt to rectify its shortcomings relies on assumptions. As the properties of the apparently SM-like Higgs couplings are not yet fully understood, hence the electroweak symmetry breaking has not been established experimentally yet, its interactions provide a viable avenue towards the discovery of new physics. A further assumption that opens up a broad range of modifications to particle physics is the heaviness of new physics relative to the scales we currently explore directly. This assumption does not give rise to an exhaustive beyond-the-SM (BSM) exploration programme as new particles can readily hide from experimental scrutiny~\cite{Gaemers:1984sj,Jung:2015gta,Basler:2019nas}. Keeping this caveat in mind, a heavy BSM spectrum enables the application of effective field theory (EFT) in a bottom-up approach~\cite{Weinberg:1978kz}; this interpretative framework is increasingly considered by the LHC experiments. As a tool that connects and mediates phenomenological observations between different scales, the value and suitability of an EFT are unquestionable. With high-precision successor experiments to the LHC being actively considered at CERN~\cite{selvaggi_2025_n2emg-43f06}, the recent progress of~\cite{Asteriadis:2024xuk,Dawson:2024pft,Bellafronte:2025jbk,Bellafronte:2026mhp} demonstrates that an $e^+e^-$ machine can probe radiative imprints of effective interactions in $e^+e^-\to hZ$ that can be contextualised within extensions of the weak sector through appropriate matching~\cite{Cohen:2020xca,Dittmaier:2021fls,Dittmaier:2026nnb}.

At very high precision, many phenomenological and conceptual aspects, accessible through motivated extensions in a top-down approach using concrete UV-complete models, also warrant investigation in parallel. In such analyses, experimental sensitivity can be interpreted straightforwardly, without relying on matching and related uncertainties~\cite{Alonso:2025jvv,Brivio:2025yrr,Assi:2026vjx,Dawson:2023ebe,Dawson:2023oce, DasBakshi:2024krs}, and characteristic interaction hierarchies can be modelled with comparably little overhead (for a review of electroweak precision calculations see, e.g.,~\cite{Denner:2019vbn}). 

The current future-collider programme is built around a dedicated Higgs run at an $e^+e^-$ machine, operating at a centre-of-mass energy of about 240 GeV, near the maximum of the $e^+e^- \to hZ$ cross section. The $hZ$ run is preceded by a precision $Z$-pole run that will both determine the electroweak input parameters precisely and constrain new physics indirectly, e.g.~\cite{Peskin:1991sw}. Potentially, the collision energies can be pushed to the top-quark threshold, but a dedicated multi-Higgs run is currently not foreseen. Within the specific limitations of the $e^+e^-\to hZ$ run, relatively few phenomenological handles are available compared to the flexibility offered by a hadron collider such as the LHC. Differential distributions may offer partial insight, and the various energy scales probed in the process yield distinct `nested' energy scales (i.e. the Higgs and $Z$ masses) that can further enhance the sensitivity to new physics (see~\cite{Bellafronte:2025jbk} for a recent EFT analysis).

In this work, we consider $e^+e^-$ collisions at $\sqrt{s}=240\ \mathrm{GeV}$ in the Two-Higgs-Doublet Model (2HDM). Although this approach is model-dependent, it nevertheless captures the theoretical handles that drive $hZ$ production beyond tree-level: (i) the presence of new states with a well-defined decoupling limit, (ii) modifications of the Higgs self-coupling from its SM expectation, and (iii) the possibility of Higgs mixing and non-standard Higgs couplings. 
Radiative corrections correlate self-interactions and Higgs mixing; this interplay is particularly interesting as $hZ$ production has been motivated as a probe of the Higgs self-interaction~\cite{McCullough:2013rea,DiVita:2017vrr}. 

This paper is structured as follows. In~\cref{sec:details}, we briefly review the 2HDM and details of the $e^+e^-\to hZ$ process to make this work self-contained. Particular emphasis is given to the alignment and decoupling limits of the model as these frame an indirect precision analysis of BSM effects at the weak scale. \cref{sec:results} is dedicated to results. We identify the dominant phenomenological drivers that can be accessed with a projected inclusive sensitivity of $\delta\sigma/\sigma \simeq 0.3\%$~\cite{selvaggi_2025_n2emg-43f06} in $e^+e^-\to hZ$; we also discuss the complementarity that can be gained by isolating exclusive Higgs boson decays in fermionic final states. We offer conclusions from our results in \cref{sec:conc}. 

%%%%%%%%%%%%%%%%%%%%%%%%%%%%%%%%%%%%%%%
\section{The Two-Higgs-Doublet Model}
\label{sec:details}
%%%%%%%%%%%%%%%%%%%%%%%%%%%%%%%%%%%%%%%
In this work, we consider the 2HDM~\cite{Lee:1973iz,Branco:2011iw,Gunion:2002zf} as a benchmark BSM model with an extended Higgs sector. The CP-conserving scalar potential with the softly broken discrete $\mathbb{Z}_{2}$ symmetry is given as
\begin{multline}
V(\Phi_1, \Phi_2) = {m}^2_{11}(\Phi_1^{\dagger}\Phi_1) + {m}^2_{22}(\Phi_2^{\dagger}\Phi_2) - {m}^2_{12}(\Phi_1^{\dagger}\Phi_2 
+ \Phi_2^{\dagger}\Phi_1) + {{\lambda}_1}(\Phi_1^{\dagger}\Phi_1)^2 + {{\lambda}_2}(\Phi_2^{\dagger}\Phi_2)^2  \\
 +  {\lambda}_3(\Phi_1^{\dagger}\Phi_1)(\Phi_2^{\dagger}\Phi_2) + {\lambda}_4(\Phi_1^{\dagger}\Phi_2)(\Phi_2^{\dagger}\Phi_1) +
\frac{1}{2}{\lambda}_5[(\Phi_1^{\dagger}\Phi_2)^2 + (\Phi_2^{\dagger}\Phi_1)^2]\,,
\end{multline}
where the doublets $\Phi_{1,2}$ are expanded as 
\begin{equation}
	\Phi_{1}= \begin{pmatrix}
	\phi^{+}_{1} \\
	\frac{1}{\sqrt{2}}(v_{1}+\zeta_{1}+i \psi_{1})
		\end{pmatrix} \,, \hspace{1.5cm} 
		\Phi_{2}=\begin{pmatrix}
		\phi^{+}_{2} \\ 
		\frac{1}{\sqrt{2}}(v_{2}+\zeta_{2}+i \psi_{2})
	\end{pmatrix} \,.
\end{equation}
The addition of the second Higgs doublet in the 2HDM gives rise to five physical Higgs bosons.
Under the assumption of CP conservation, these states are two neutral CP-even Higgs bosons, $h$ and $H$ (where, by convention, $m_h<m_H$), one neutral CP-odd Higgs boson, $A$, and two charged Higgs bosons, $H^\pm$. Throughout this work, we will identify the observed Higgs boson with the light CP-even state, such that $m_h=125\ \mathrm{GeV}$. We define the mixing angles $\alpha$ and $\beta$ such that they diagonalise the CP-even sector and the CP-odd and charged sectors of the scalar potential, respectively. Furthermore, $\tb$ can be related to the ratio of the two vacuum expectation values (vevs) of the two Higgs doublets as 
\begin{align}
\tan\beta = \frac{v_2}{v_1} \,.
\end{align}
The combination of these two mixing angles controls the modifications of the tree-level Higgs couplings within the 2HDM with respect to the corresponding couplings of the SM Higgs boson. For the tree-level couplings of the neutral Higgs bosons to massive gauge bosons, the ratios to the corresponding SM predictions are given by
\begin{equation}
    \zeta_V^h=\sbma\,,\quad \zeta_V^H=\cbma\,,\quad \zeta_V^A=0\,,
    \label{eq:zeta-V}
\end{equation}
with $V=W^\pm,Z$, where we use the shorthand notation $s_x\equiv \sin(x)$ and $c_x\equiv \cos(x)$.
For later convenience and comparability with the existing literature, we also define $\kappa_V=\zeta_V^h$.
For the definition of quartic couplings between two Higgs bosons and two gauge bosons, as well as for the trilinear and quartic Higgs self-couplings, we refer to \cite{Gunion:2002zf,ArcoGarcia:2023zjz}.

To avoid flavour-changing neutral currents at the tree level, it is common to impose a discrete $\mathbb{Z}_2$ symmetry under which the two Higgs doublets have different parities~\cite{Glashow:1976nt,Paschos:1976ay}. When promoting this symmetry to the Yukawa interactions, only four non-equivalent Yukawa textures are possible, which are known as the four 2HDM Yukawa types~\cite{Aoki:2009ha}. In this work, we will consider only the type-I 2HDM, where the fermion doublets couple only to one Higgs doublet. Nevertheless, the Yukawa interactions do not play a significant role in the results of this work. The tree-level couplings between fermions and the neutral Higgs bosons are again controlled by the angles $\alpha$ and $\beta$, where the ratios to the corresponding SM predictions in type I are given by
\begin{equation}
    \zeta_f^h = \sbma+\cbma\,\cot\beta \,,\quad 
    \zeta_f^H = \cbma-\sbma\,\cot\beta \,,\quad 
    \zeta_{u}^A=-\zeta_{d,l}^A= \cot\beta \,,
    \label{eq:zeta-f}
\end{equation}
where an additional $\gamma^5$ is required in the pseudoscalar couplings of $A$. The introduced $\mathbb{Z}_2$ symmetry can be softly broken by a dimension-two mass parameter denoted as $m_{12}^2$, and for convenience, we define the mass parameter 
\begin{equation}
\mbar\equiv\sqrt{\frac{m_{12}^2}{\sin\beta\cos\beta}}\,.
\end{equation} 
Consequently, making use of the minimisation conditions of the potential and requiring that the Higgs vevs add up to the SM vev $v\simeq 246.22\ \mathrm{GeV}$, the tree-level Higgs sector of the 2HDM can be completely determined with the following seven additional physical input parameters: 
\begin{equation}
\label{eq:parameters}
    m_h\,,\quad m_H\,,\quad m_A\,,\quad m_{H^\pm}\,,\quad \tb\,,\quad \cbma\,,\quad \mbar\,.
\end{equation}

The addition of the second doublet introduces new triple and quartic interactions among the scalar states of the 2HDM, including the physical Higgs bosons and the Goldstone bosons. Of particular importance is the triple Higgs self-coupling of $h$, whose ratio to the SM value is given by
\begin{equation}
    \kala = \sbma^3+\sbma\cbma^2\left(3-2\frac{\mbar^2}{\mh^2}\right)+2\cbma^3\cot2\beta\left(1-\frac{\mbar^2}{\mh^2}\right)\,.
    \label{eq:kala}
\end{equation}

%%%%%%%%%%%%%%%%%%%%%%%%%%%%%%%%%%%%%%%
\subsection{Constraints on the 2HDM}
\label{sec:constraints}
%%%%%%%%%%%%%%%%%%%%%%%%%%%%%%%%%%%%%%%
In this section, we briefly summarise the main theoretical and experimental constraints on the 2HDM that are considered in this work, as implemented in the public code {\tt ScannerS}~\cite{Coimbra:2013qq,Muhlleitner:2020wwk}: 
\begin{cedescription}
    \item[Electroweak precision data:]
    A common way to parametrise BSM physics in electroweak (EW) precision observables is via the oblique parameters $S$, $T$ and $U$~\cite{Peskin:1990zt}.
    We use the $2\sigma$ allowed region given by a $\chi^2$ fit to the reported values in~\cite{Haller:2018nnx}. The most constraining parameter is $T$, which can receive large corrections in the 2HDM. This requires one neutral Higgs boson to be nearly degenerate with the charged Higgs boson~\cite{Funk:2011ad}.
    \item[Tree-level perturbative unitarity:]
    We require that the eigenvalues of the tree-level $s$-wave $2\to2$ scalar scattering matrix in the high-energy limit respect unitarity bounds. For the explicit expressions in terms of the quartic couplings of the potential, see~\cite{Akeroyd:2000wc,Ginzburg:2005dt}.
    \item[Potential stability:]
    We require the tree-level potential to be bounded from below~\cite{Deshpande:1977rw}. Additionally, we require that the EW minimum is a global minimum of the potential~\cite{Barroso:2013awa}.
    \item[BSM Higgs boson searches:]
    We take into account the latest 95\% CL experimental bounds from searches for BSM Higgs bosons carried out at the LHC, as implemented in the public code {\tt HiggsBounds}~\cite{Bechtle:2008jh,Bechtle:2011sb,Bechtle:2013wla,Bechtle:2015pma,Bechtle:2020pkv}, as part of {\tt HiggsTools}~\cite{Bahl:2022igd}. 
    \item[Signal strength measurements for the SM-like Higgs boson:]
    We require the properties of the lightest CP-even Higgs boson $h$ to be in agreement with the rate measurements of the discovered Higgs boson, as given by a $\chi^2$ fit developed in the public code {\tt HiggsSignals}~\cite{Bechtle:2013xfa,Bechtle:2014ewa,Bechtle:2020uwn}, as part of {\tt HiggsTools}~\cite{Bahl:2022igd}. Concretely, we allow for a $2\sigma$ (i.e.~$\Delta\chi^2\leq6.18$) deviation from the SM fit value.
    \item[Flavour observables:]
    The presence of the charged Higgs boson can give sizeable contributions to some $B$-meson decays, such as $B\to X_s\gamma$ or $B_s\to\mu\mu$. To take them into account, we use the results of \cite{Haller:2018nnx} at the $2\sigma$ level in the $m_{H^\pm}$--$\tb$ plane.
\end{cedescription}

%%%%%%%%%%%%%%%%%%%%%%%%%%%%%%%%%%%%%%%
\subsection{Decoupling Limit vs. Alignment Limit}
\label{sec:dec-vs-align}
%%%%%%%%%%%%%%%%%%%%%%%%%%%%%%%%%%%%%%%
Given the absence of a clear sign of BSM physics in the Higgs boson sector at the moment, together with the unprecedented precision expected at future colliders, it is pertinent to discuss under which conditions the predictions of the 2HDM approach those of the SM. One possibility to achieve this is the so-called {\it decoupling limit} (or the heavy mass limit) of the 2HDM, which is defined as the limit in which all Higgs bosons become very heavy, except for the light CP-even Higgs boson $h$, while the quartic couplings remain perturbative and respect unitarity~\cite{Gunion:2002zf}. In general, the Higgs bosons can receive their masses from the soft-breaking mass term and from the EW spontaneous symmetry breaking, which can be schematically written as $m_\phi^2\sim \mbar^2+\lambda_\phi v^2$ for a Higgs boson $\phi$, where $\lambda_\phi$ is a combination of quartic couplings from the potential. Therefore, in the regime $\mbar^2 \gg \lambda_\phi v^2 $, heavy Higgs boson masses imply large values for the soft-breaking mass parameter.\footnote{Notice that it is not possible to define the decoupling limit in a 2HDM with an exact $\mathbb{Z}_{2}$ symmetry, that is with $m_{12}^2=\mbar=0$, given the fact that the theory becomes non-unitary for Higgs boson masses $\gtrsim 700\ \mathrm{GeV}$~\cite{Kanemura:2004mg,Akeroyd:2000wc}.} Consequently, the decoupling limit can be defined by
\begin{equation}
    \mH\,,\mA\,,\mHp\,,\mbar\sim \Lambda_\mathrm{2HDM} \gg v\,.
    \label{eq:decoupling}
\end{equation}
Within this limit, the tree-level couplings of $h$ to gauge bosons and fermions, as well as the $h$ triple and quartic self-couplings, tend to their SM values, because these conditions amount to~\cite{Gunion:2002zf,Dawson:2023ebe}
\begin{equation}
    \cbma\sim\left(\frac{v}{\Lambda_{\mathrm{2HDM}}}\right)^2\to0\,,
\end{equation}
as can be seen in~\cref{eq:zeta-V,eq:zeta-f,eq:kala}.

The discussion has revolved so far around tree-level quantities, but the decoupling limit defined in \cref{eq:decoupling} also implies that the 2HDM predictions approach those of the SM at higher orders, following the Appelquist-Carazzone theorem~\cite{Appelquist:1974tg}. For this result, the assumption of perturbative couplings is essential, since it implies that the heavy Higgs bosons receive their masses mainly from the soft-breaking mass parameter $\mbar$, rather than from the EW symmetry breaking mechanism.\footnote{The difference in the origin of the mass terms for the heavy particles leaves footprints in the phenomenology \cite{Georgi:1977gs,Crawford:2024nun,Kilic:2026ogm}, even in EFT descriptions \cite{Cohen:2020xca,Banta:2021dek,Asiain:2026sio}.}

Another possibility to recover the tree-level SM prediction in the 2HDM is to directly set
\begin{equation}
    \cbma\simeq 0\,,
\end{equation}
regardless of any further assumptions about the Higgs boson masses, since it is a free parameter of the theory. This is known as the {\it alignment limit} and, as discussed above, it predicts SM-like couplings for the Higgs boson $h$ at tree level.\footnote{A symmetry-based argument for alignment is provided in Ref.~\cite{BhupalDev:2014bir}} Nevertheless, in the case of alignment without decoupling, higher-order corrections in the 2HDM can differ from their SM counterparts, even with only SM particles in external legs.

In particular, for $hZ$ production at $e^+e^-$ colliders, the 2HDM scalar sector plays an essential role in the cross section prediction. As we will discuss in \cref{sec:results}, one-loop effects mediated by BSM Higgs bosons can produce deviations from the SM prediction, even if the alignment limit is imposed. For brevity, we provide here the expressions of the triple Higgs couplings that appear in the one-loop prediction of the Higgsstrahlung cross section in the alignment limit ($\cbma=0$):\footnote{A complete set of Feynman rules of the 2HDM can be found in appendix A of Ref.~\cite{ArcoGarcia:2023zjz}.
}
\begin{equation}
    \begin{aligned}
        \lambda_{hhh}& = \frac{3m_h^2}{v^2}\,, \\
        \lambda_{hhH}& = 0\,, \\
        \lambda_{h\phi\phi} & = \frac{m_h^2+2 m_\phi^2-2 \bar{m}^2}{v^2}\,,\\
    \end{aligned}
    \label{eq:thc}
\end{equation}
where $\phi= H, A,  H^\pm$. Notice that, for degenerate masses, the scalar coupling $\lahHH$ is only equal to $\lahAA$ and $\lahHpHp$ in the alignment limit, while $\lahAA=\lahHpHp$ always holds. We define these Higgs couplings through the terms  $V \supset \lambda_{h\phi\phi^\prime}\,v\,h\phi\phi^\prime$ in the 2HDM potential. Note that, in the decoupling limit, these scalar couplings of $h$ to heavy Higgs bosons remain small, of order $\sim m_h^2/v^2$, such that the effect of heavy degrees of freedom decouples as they become heavy. In contrast, if some BSM Higgs bosons remain relatively light and a splitting is introduced between the physical masses and the soft-breaking mass $\mbar$, those triple Higgs couplings can be large, even within the requirement of perturbative unitarity~\cite{Arco:2020ucn,Arco:2022xum}. This particular region of the parameter space will be of great interest in the present work because large scalar couplings can lead to sizeable deviations in the $hZ$ cross section.

%%%%%%%%%%%%%%%%%%%%%%%%%%%%%%%%%%%%%%%
\section{Results}
\label{sec:results}
%%%%%%%%%%%%%%%%%%%%%%%%%%%%%%%%%%%%%%%
First, we briefly summarise the setup of our computation, with further details given in Ref.~\cite{Anisha:2025zbc}. We perform the analytic computation using {\tt{FeynArts}}, {\tt{FormCalc}}/{\tt{FeynCalc}}, and {\tt{LoopTools}}~\cite{vanOldenborgh:1989wn,Mertig:1990an,Hahn:2000kx,Hahn:1998yk,Hahn:2000jm,Shtabovenko:2016sxi,Shtabovenko:2020gxv}. In particular, we use the built-in {\tt{FeynArts}} model file {\tt{THDM.mod}}, based on Ref.~\cite{Gunion:1989we}.

We retain a non-zero electron mass to regularise collinear divergences in the QED corrections to off-shell $Z$ production. The remaining IR divergences are regularised by including soft-photon emission up to $E_\gamma = 30$ GeV. We perform the rest of our computation with a massless electron, which allows us to neglect the Yukawa-like couplings. 

We employ the alternative Fleischer-Jegerlehner tadpole scheme~\cite{Fleischer:1980ub}, in which the additional renormalisation condition of vanishing tadpoles is imposed. In the context of the 2HDM, such a scheme is also referred to as `alternative tadpole scheme'~\cite{Krause:2016oke,Krause:2016xku,Krause:2017mal,Krause:2018wmo}. Additional discussions on the renormalisation of the 2HDM are given in Refs.~\cite{Kanemura:2004mg,Denner:2016etu,Altenkamp:2017ldc,Altenkamp:2017kxk,Fox:2017hbw,Grimus:2018rte,Denner:2018opp,Dittmaier:2022maf,Dittmaier:2022ivi,Kanemura:2024ium,Guerandel:2025kjq}. 

%%%%%%%%%%%%%%%%%%%%%%%%%%%%%%%%%%%%%%%
\subsection{Anatomy of the Cross Section}
\label{sec:anatomy}
%%%%%%%%%%%%%%%%%%%%%%%%%%%%%%%%%%%%%%%
%%%%%%%%%%%%%%%%%%%%%%%%%%%%%%%%%%%%%%%
\begin{figure}[t]
    \centering
    \includegraphics[scale=0.9]{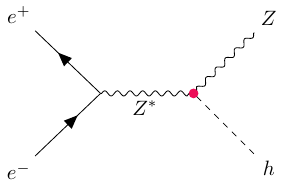}
    \caption{Leading-order Feynman diagram contributing to the Higgsstrahlung cross section.}
    \label{fig:FD-LO}
\end{figure}
%%%%%%%%%%%%%%%%%%%%%%%%%%%%%%%%%%%%%%%
%%%%%%%%%%%%%%%%%%%%%%%%%%%%%%%%%%%%%%%
\begin{figure}[!b]
    \centering
    \subfigure[]{\includegraphics[scale=0.7]{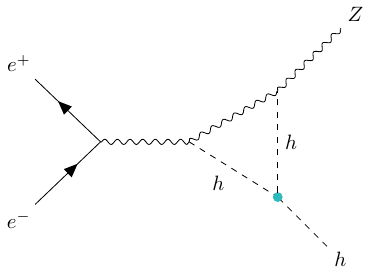}}\hfill
    \subfigure[]{\includegraphics[scale=0.7]{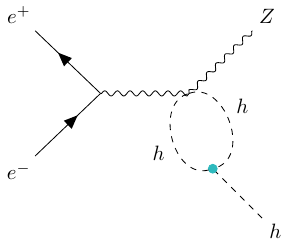}}\hfill
    \subfigure[]{\includegraphics[scale=0.7]{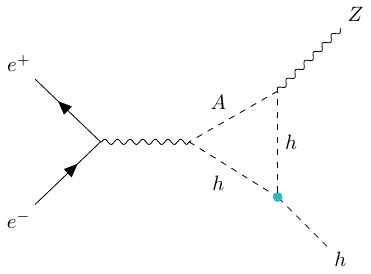}}
    \parbox{0.25\textwidth}{\centering
    \subfigure[]{\includegraphics[scale=0.7]{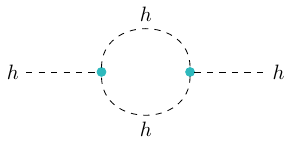}}}
    \hfill
    \parbox{0.7\textwidth}{
    \caption{Feynman diagrams sensitive to $\kala$ in the 2HDM.
    Diagrams (a), (b) and (d) contribute to the cross section as given by the factor of $\delta_\lambda$ in \cref{eq:kappa}.
    Diagram (c) is not included in \cref{eq:kappa}, and it is suppressed by a factor of $\cbma^2$.
    These and all other relevant diagrams are included in our full computation of the cross section.}
    \label{fig:FD-kala}}
\end{figure}
%%%%%%%%%%%%%%%%%%%%%%%%%%%%%%%%%%%%%%%

In extensions of the SM, radiatively generated coupling deviations from the SM limit have been studied extensively~\cite{Englert:2013tya,Craig:2013xia} and can provide an avenue for new physics discovery~\cite{FCC:2018byv}. By the same token, modifications of the SM Higgs boson's self-interactions can be mapped onto a coupling modification that yields the associated production cross section at lepton colliders~\cite{McCullough:2013rea,DiVita:2017vrr}\footnote{This equation is valid for $\sqrt{s}=240\ \mathrm{GeV}$ since the loop factor associated with $\delta_\lambda$ is energy dependent.}
\begin{equation}
\label{eq:kappa}
\frac{\sigma_{\mathrm{eff}}}{\sigma_{\mathrm{SM}}}\sim 1+2\delta_Z+0.014\delta_\lambda
\equiv1+\delta_{\rm eff}
\,,
\end{equation}
where $\delta_Z=\kappa_Z-1$ is the modification of the tree-level coupling of the Higgs boson to $Z$ bosons, which enters at leading order in the diagram of \cref{fig:FD-LO}. $\delta_\lambda=\kappa_\lambda-1$ denotes the departure of the Higgs self-coupling from its SM value, reflecting the relative loop suppression. This expression will help us estimate the effects of mixing and the Higgs self-coupling on the Higgsstrahlung cross section.
Note that the $\kala$ factor in \cref{eq:kappa} accounts for the SM-like contributions from one-loop diagrams involving SM particles and the renormalised Higgs boson wavefunction. 
These contributions are represented by diagrams (a), (b) and (d) of \cref{fig:FD-kala}.
In the 2HDM, there is an additional contribution involving the pseudoscalar boson $A$, which depends on $\kala$, shown in \cref{fig:FD-kala}(c).
However, this diagram is suppressed by a factor of $\cbma^2$, and therefore \cref{eq:kappa} remains a good approximation to the effect of $\kala$ in the cross section in the 2HDM.

The coupling modifier $\kappa_Z$ in the 2HDM is given by $\kappa_Z=\zeta_V^h=s_{\beta-\alpha}$, see \cref{eq:zeta-V}, while the self-coupling modifier $\kala$ is given in \cref{eq:kala}. 
Therefore, as the SM alignment limit, $\cos(\beta-\alpha)\simeq \epsilon$, is approached at tree level, the relevant 2HDM quantities scale as
\begin{equation}
\label{eq:decoupl}
\frac{\delta_\lambda}{2\delta_Z} \simeq  \frac{2 \bar m^2}{m_h^2} -\frac{3}{2} +  O(\epsilon)\,,
\end{equation}
with $\delta_Z,\delta_\lambda=O(\epsilon^2)$.\footnote{There is a non-perturbative region in $\tan\beta$ where the suppression can be ameliorated through the $\tan\beta$-enhanced ${\cal{O}}(\epsilon)$ contribution in~\cref{eq:decoupl}. This is in tension with theoretical constraints, and we will not consider this region further.}  Therefore, for values around $\mbar \simeq 750\ \mathrm{GeV}$, the loop suppression can naively be compensated by the enhanced Higgs self-coupling in Eq.~\eqref{eq:kappa}. This significant self-coupling modification requires $\bar m \gg m_h$, which also sets the mass scale of the exotics associated with the enlarged Higgs sector. However, a large value of $\bar m$ implies alignment through decoupling, as discussed in \cref{sec:dec-vs-align}, and consequently the cross section will approach its SM value. 
Therefore, deviations from the SM induced by $\delta_\lambda$ require (i) misalignment, which implies that deviations induced by the $\delta_Z$ mixing will also be present, and (ii) TeV-scale values of $\mbar$, which set the scale for the new scalars. 

%%%%%%%%%%%%%%%%%%%%%%%%%%%%%%%%%%%%%%%
\begin{figure}[t]
    \centering
    \subfigure[]{\includegraphics[scale=0.7]{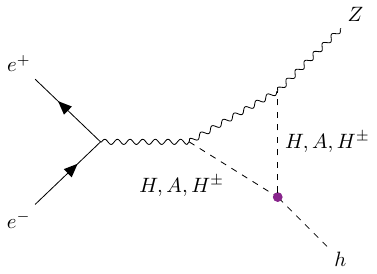}}\hfill
    \subfigure[]{\includegraphics[scale=0.7]{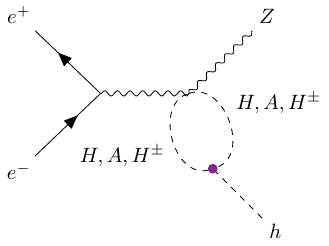}}\hfill
    \subfigure[]{\includegraphics[scale=0.7]{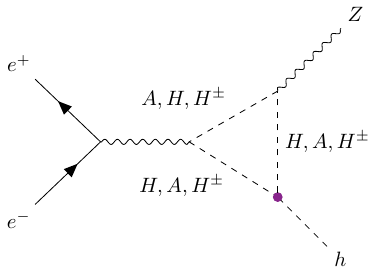}}\\
\parbox{0.25\textwidth}{\subfigure[]{ \includegraphics[scale=0.7]{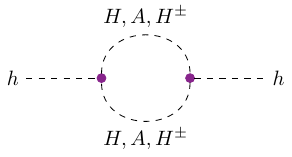}}}\hfill
\parbox{0.7\textwidth}{\caption{Feynman diagrams with additional non-SM-like scalars whose contribution to the Higgsstrahlung cross section is non-zero in the alignment limit. Note that these diagrams are not sensitive to $\lambda_{hhh}$ but to $\lambda_{h\phi \phi}$.}
    \label{fig:FD-scalars}}
\end{figure}
%%%%%%%%%%%%%%%%%%%%%%%%%%%%%%%%%%%%%%%

We can estimate the contributions of these non-SM-like states by comparing our cross section prediction with the effective result of \cref{eq:kappa}, which only accounts for the dominant effects induced by $\delta_Z$ and $\delta_\lambda$:
\begin{equation}
    \delta_{\mathrm{scalars}} \simeq \delta_{\mathrm{2HDM}} - \delta_{\mathrm{eff}}\,.
    \label{eq:scalars}
\end{equation}
It is important to highlight that these kinds of BSM contributions do not necessarily vanish in the alignment limit, i.e.~$\cos(\beta-\alpha)=0$. Therefore, in the 2HDM, it is possible to induce a deviation from the SM value even if the tree-level $hZZ$ and $hhh$ couplings take their SM values. In particular, in the exact alignment limit, {\emph{only}} the additional Higgs bosons can produce a deviation in the prediction 
\begin{equation}
    \delta_{\rm scalars}^{\rm align} = \delta_{\rm 2HDM}^{\rm align}\,.
    \label{eq:align}
\end{equation}
The set of Feynman diagrams that do not vanish in the alignment limit is shown in~\cref{fig:FD-scalars}. We stress that all these contributions depend on one triple coupling between one SM-like Higgs boson and two extra Higgs bosons, namely $\lahHH$, $\lahAA$ or $\lahHpHp$.

%%%%%%%%%%%%%%%%%%%%%%%%%%%%%%%%%%%%%%%
\subsection{Parameter Scan in the 2HDM }
%%%%%%%%%%%%%%%%%%%%%%%%%%%%%%%%%%%%%%%
As an initial step, we explore the 2HDM prediction for the $Z$-associated Higgs production by means of a parameter scan, where we allow the free parameters of \cref{eq:parameters} to vary within the following intervals:%
\footnote{This corresponds to the same intervals considered in our previous work~\cite{Anisha:2025zbc}.}
\begin{equation}
\begin{gathered}
m_H \in \left[150,\, 1500\right] \ \mathrm{GeV}\,, \quad 
    m_A,\ m_{H^\pm} \in \left[20,\, 1500\right] \ \mathrm{GeV}\,, \quad \\
    \tan\beta \in \left[0.5,\, 50\right]\,, \quad 
    \cos\!\left(\beta-\alpha\right) \in \left[-0.35,\, 0.35\right]\,, \quad 
    \bar m \in \left[0,\, 1500\right] \mathrm{GeV} \,,
\end{gathered}
\label{eq:scan}
\end{equation}
and we require them to satisfy the 2HDM constraints discussed in \cref{sec:constraints}.

%%%%%%%%%%%%%%%%%%%%%%%%%%%%%%%%%%%%%%%
\begin{figure}[!t]
    \centering
    \includegraphics[width=0.49\linewidth]{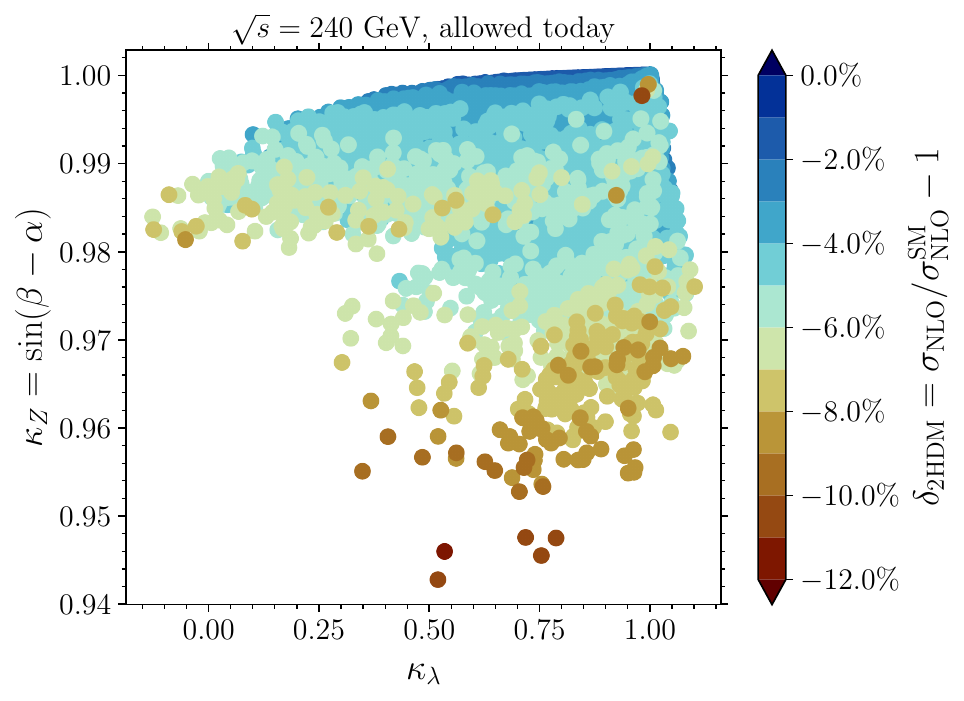}
    \includegraphics[width=0.49\linewidth]{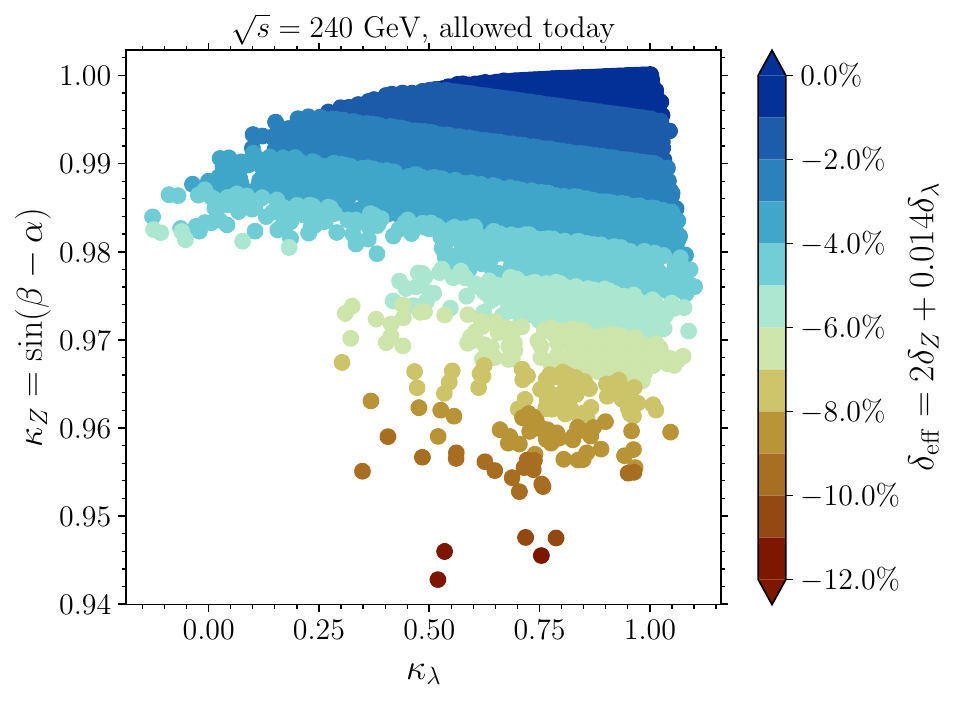}
    \includegraphics[width=0.49\linewidth]{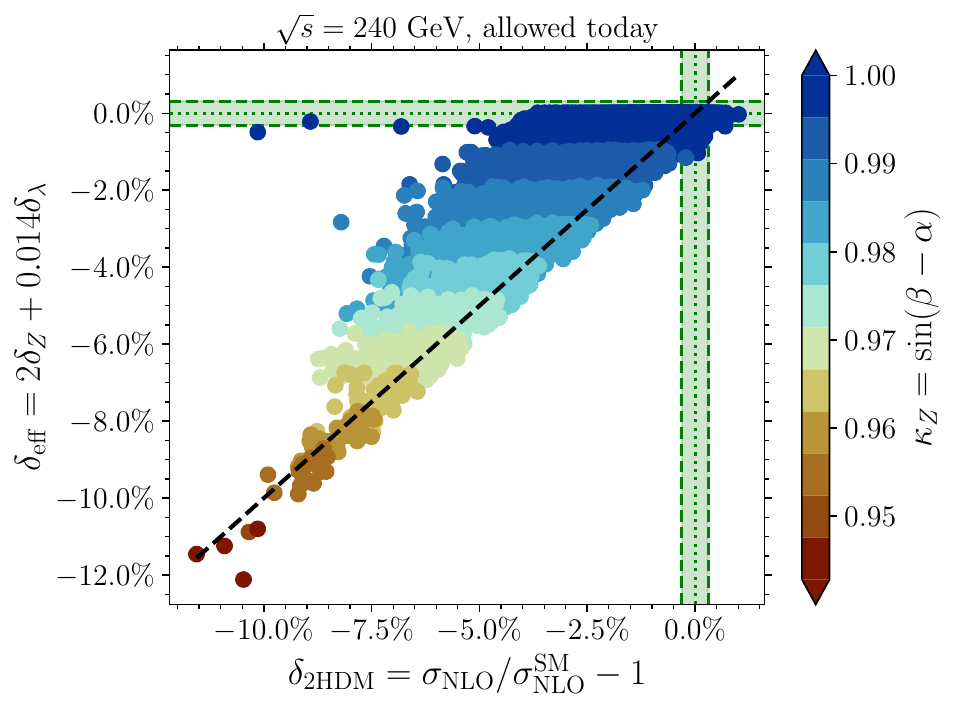}
    \includegraphics[width=0.49\linewidth]{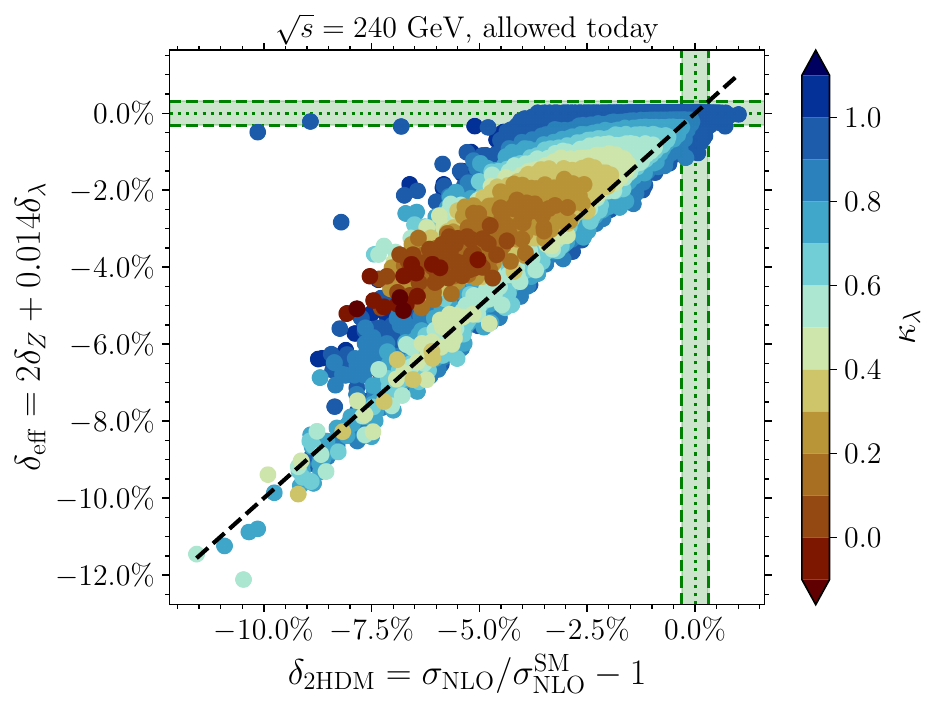}
    \caption{Higgsstrahlung cross section in the 2HDM compared with the expected result from only non-SM values of $\kappa_Z=1+\delta_Z$ and $\kala=1+\delta_\lambda$ at $\sqrt{s}=240\ \mathrm{GeV}$ for the allowed points in the parameter scan of \cref{eq:scan}. 
    The green bars denote the expected experimental uncertainty at FCC-ee of 0.31\%~\cite{selvaggi_2025_n2emg-43f06}, and the black dotted line denotes $\delta_\mathrm{2HDM}=\delta_\mathrm{eff}$.}
    \label{fig:EffvsNLO}
\end{figure}
%%%%%%%%%%%%%%%%%%%%%%%%%%%%%%%%%%%%%%%

We show the results of the $Z$-associated Higgs production cross section for the allowed points of the scan in the left plots of \cref{fig:EffvsNLO}. We generally find negative deviations with respect to the SM prediction,\footnote{Our SM reference value for the NLO cross section is $\sigma_{\text{NLO}}^{\text{SM}} = 0.211\ \mathrm{pb}$, as in our previous work~\cite{Anisha:2025zbc}. We obtain this value in a decoupling scenario of the 2HDM, with $\mH=\mA=\mHp=\mbar=1.8\ \mathrm{TeV}$, $\cbma=0$ and $\tan\beta=2$.} reaching values of about $-10\%$. The largest deviations are mainly associated with large distortions in $\kappa_Z$, since this parameter already enters the tree-level prediction. For comparison, we show in the right plots the expected deviation in the cross section considering only BSM effects from a non-SM value of $\kappa_Z$ and/or $\kala$, as given by $\delta_\mathrm{eff}$ in \cref{eq:kappa}. It can be seen that our prediction for the cross section in the 2HDM is qualitatively different from $\delta_\mathrm{eff}$. In particular, the largest differences between the two predictions occur in the alignment limit (implying $\kappa_Z=\kala=1$), while they agree more closely in misaligned scenarios with $\kappa_Z\lesssim0.96$.

This comparison points to the fact that there are other BSM effects contributing to the 2HDM prediction beyond the `na\"ive' effects from a non-SM value of $\kappa_Z$ and $\kala$, particularly those arising from the additional Higgs bosons present in the 2HDM. Furthermore, with the presence of these exotic states, the dependence of the cross section on $\kala$ is less transparent than the $\delta_\mathrm{eff}$ prediction, which would make the extraction of the Higgs boson self-coupling in this process much more challenging. In the following sections, we will further explore these differences in concrete scenarios in the 2HDM, and we will characterise the effects of these additional Higgs bosons in the cross section.

%%%%%%%%%%%%%%%%%%%%%%%%%%%%%%%%%%%%%%%
\subsection{Misalignment and Light New Physics}
\begin{figure}[!t]
    \centering
    \includegraphics[width=0.99\linewidth]{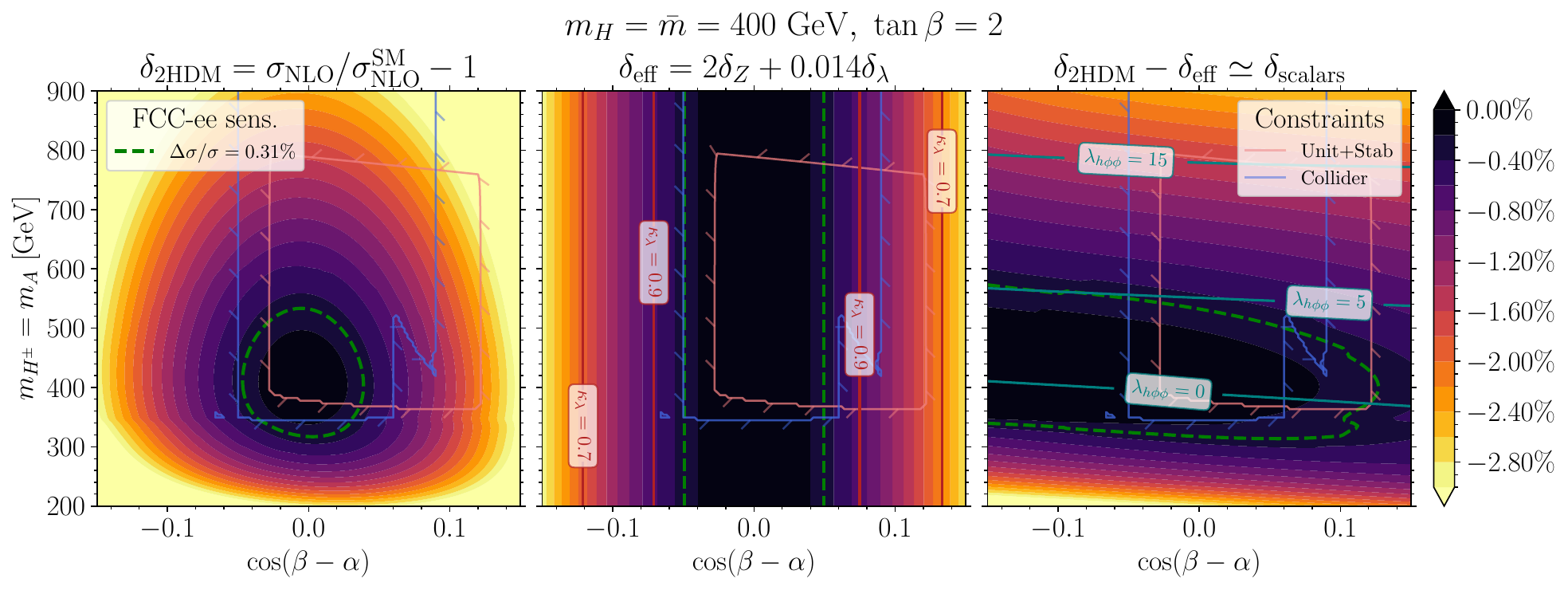}
    \caption{Higgsstrahlung cross section at $\sqrt{s}=240\ \mathrm{GeV}$ at NLO in the 2HDM. \textit{Left:} Full cross section result. \textit{Middle:} Expected result with distortions from only $\delta_Z=\kappa_Z-1$ and $\delta_\lambda=\kala-1$. \textit{Right:} Difference between the two predictions as a proxy of the contributions from extra Higgs bosons. We use the notation $\lambda_{h\phi\phi}\equiv\lambda_{hAA}=\lambda_{hH^+H^-}$.}
    \label{fig:misalignment}
\end{figure}

%%%%%%%%%%%%%%%%%%%%%%%%%%%%%%%%%%%%%%%
We start our analysis of the Higgsstrahlung cross section in a 2HDM scenario that allows $\cbma$ values away from the alignment limit, together with a light non-SM-like scalar $H$ around 400 GeV. To disentangle different contributions to the cross section, we present the results in a three-panel plot, as shown in \cref{fig:misalignment}. The left plot shows the difference between the 2HDM and the SM predictions of the Higgsstrahlung cross section at NLO, normalised to the SM prediction (i.e. the 2HDM in the decoupling limit), and the middle plot shows the cross section deviations as given by $\delta_{\mathrm{eff}}$, defined in \cref{eq:kappa}. The right plot shows the difference between the first two, denoted by $\delta_{\mathrm{scalars}}$ as defined in \cref{eq:scalars}. Consequently, the middle plot shows the main effects from the tree-level Higgs coupling to $Z$ bosons and from the Higgs self-coupling, while the right plot serves as an approximation to the contributions from the non-SM diagrams mediated by additional BSM Higgs bosons. In addition, in the middle and right panels, we show the triple Higgs coupling relevant to the corresponding contribution, i.e. $\kala$ in $\delta_\mathrm{eff}$, shown in dark red, and $\laphi \equiv \lambda_{hAA} = \lambda_{hH^+H^-}$ in $\delta_\mathrm{scalars}$, shown in teal colour.\footnote{The other relevant coupling $\lahHH$ is not shown in our plots because it is always small compared to $\lahAA$ and $\lahHpHp$ due to the choice of $m_H\sim\mbar$.} We also show the region allowed by the unitarity and stability constraints (enclosed by the pink contours) and the region allowed by collider searches and measurements of the SM-like Higgs boson (enclosed by the blue contours). All parameter-space regions shown are allowed by flavour constraints.

%%%%%%%%%%%%%%%%%%%%%%%%%%%%%%%%%%%%%%%
%%%%%%%%%%%%%%%%%%%%%%%%%%%%%%%%%%%%%%%

The left plot in \cref{fig:misalignment} shows the predicted deviations from the SM in the Higgsstrahlung cross section in the $\cbma$ vs.~$m_A=m_{H^\pm}$ plane, while we fix $m_H=\mbar$ and $\tan\beta=2$. It can be seen that the (always negative) deviations from the SM increase with increasing values of $\left|\cbma\right|$ and an increasing splitting between $m_A=m_{H^\pm}$ and $m_H=\mbar$. The first effect can be traced to Higgs mixing and the Higgs triple self-coupling, as seen in the prediction of $\delta_\mathrm{eff}$ in the middle plot, where both effects are driven by the departure from the alignment limit. The second effect arises from additional loop contributions mediated by heavier Higgs bosons, as reflected by the difference $\delta_\mathrm{2HDM}-\delta_\mathrm{eff}\simeq\delta_\mathrm{scalars}$ in the right plot. Large differences between $m_A,\ m_{H^\pm}$ and $\mbar$ lead to large values of the triple Higgs couplings $\lahAA,\ \lahHpHp$ (see~\cref{eq:thc}), enhancing the contributions from loop diagrams mediated by these Higgs bosons. It is worth highlighting that the effects from BSM scalars may become relevant even if they are heavy, as can be inferred from \cref{fig:misalignment} (right).

These results demonstrate that deviations in the Higgsstrahlung cross section always appear through the {\emph{combination}} of Higgs mixing and Higgs sector self-couplings. In particular, the contributions of the heavy Higgs states 
can compete with the effects induced by modifications of the 125 GeV Higgs boson self-coupling. These results are specific to the 2HDM and the way in which the decoupling and alignment limits translate to the model's parameter space. Nevertheless, they provide an example of a scenario where the interpretation of the associated-production cross section in terms of the underlying model cannot be understood solely in terms of $\kappa_\lambda$.

%%%%%%%%%%%%%%%%%%%%%%%%%%%%%%%%%%%%%%%
\subsection{BSM Scalars in the Alignment Limit and the Decoupling Limit}
\label{sec:prod_alignment}
%%%%%%%%%%%%%%%%%%%%%%%%%%%%%%%%%%%%%%%
%%%%%%%%%%%%%%%%%%%%%%%%%%%%%%%%%%%%%%%
\begin{figure}
    \centering
    \includegraphics[width=0.485\linewidth]{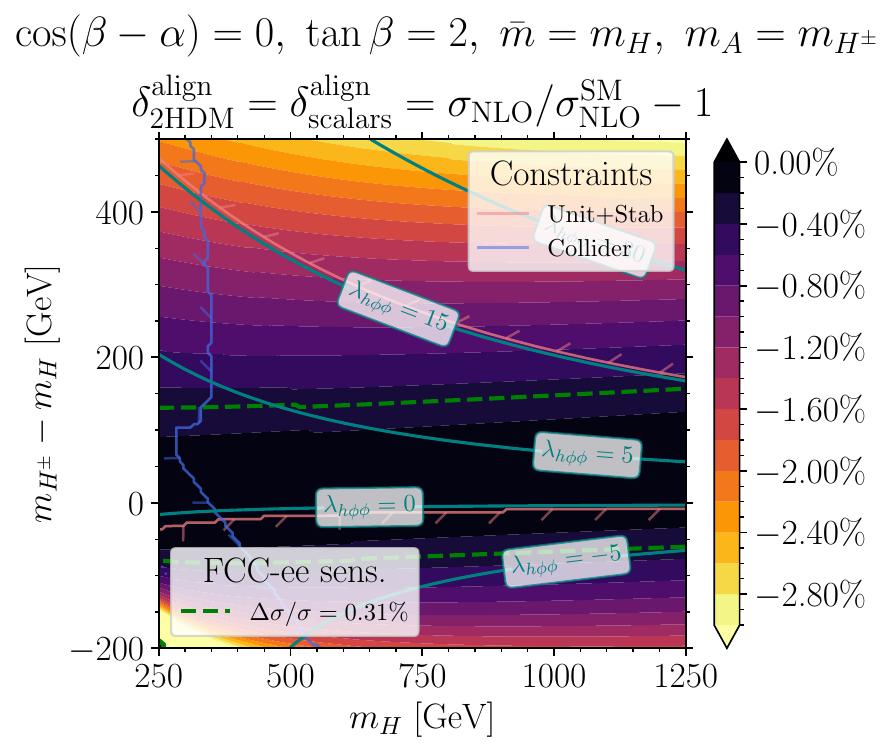}     
    \includegraphics[width=0.485\linewidth]{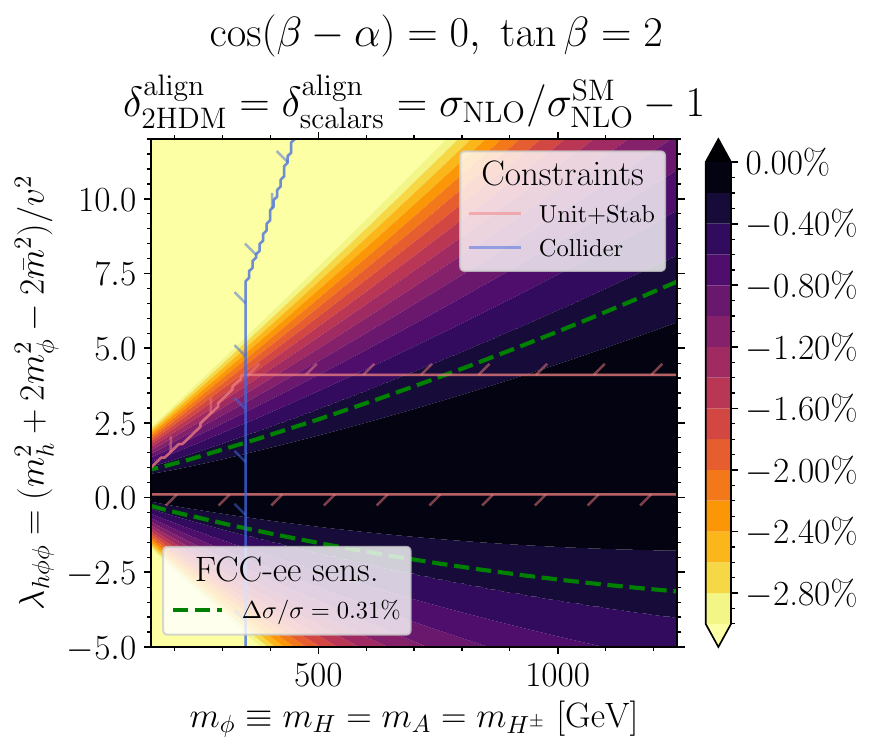} 
    \caption{Higgsstrahlung cross section at $\sqrt{s}=240\ \mathrm{GeV}$ at NLO in the 2HDM in the exact alignment limit, i.e.~$\cos(\beta-\alpha)=0$.
    Under these assumptions, only additional BSM Higgs bosons can produce a deviation in the cross section prediction, and therefore, $\delta_{\rm 2HDM}^{\rm align}=\delta_{\rm scalars}^{\rm align}$.
    }
    \label{fig:heavyHiggs}
\end{figure}
%%%%%%%%%%%%%%%%%%%%%%%%%%%%%%%%%%%%%%%

In this section, we further explore the potential deviations at NLO in the exact alignment limit, i.e.~$\cbma=0$. As discussed in \cref{sec:anatomy}, under this assumption the only possible BSM effects in the cross section are induced by one-loop diagrams mediated by additional Higgs bosons, displayed in \cref{fig:FD-scalars}.

To this end, we show in \cref{fig:heavyHiggs} the NLO Higgsstrahlung cross section in the alignment limit, and use the same colour-coding as in \cref{fig:misalignment}. In the left plot, the deviation $\delta_{\rm 2HDM}^{\rm align}$, defined in \cref{eq:align}, is shown in the plane $\mH$ vs.~$\mHp-\mH$ where we set $\mbar=\mH$, $\mA=\mHp$, and $\tan\beta=2$. The iso-cross section lines are mainly flat in the $\mH$ direction. However, the region allowed by theoretical constraints, primarily unitarity, is reduced to the fully degenerate scenario for large $\mH$ values, corresponding to the decoupling limit of the 2HDM (see discussion in \cref{sec:dec-vs-align}). Nevertheless, over the entire allowed range of $\mH$ considered in the plot, these exotic Higgs bosons can induce a deviation in the cross section beyond the projected FCC-ee sensitivity of 0.31\%~\cite{selvaggi_2025_n2emg-43f06}, which is shown by the green dashed contour. Notably, for $\mH\simeq400\ \mathrm{GeV}$ the deviation coming from only these BSM scalars can reach values up to $\delta_{\rm 2HDM}^{\rm align}\simeq-1.8\%$ when $\mHp-\mH\simeq400\ \mathrm{GeV}$. This highlights the FCC-ee's potential to probe new physics in the Higgsstrahlung channel owing to the outstanding expected experimental precision. We stress, however, that this projected precision neglects theoretical uncertainties, i.e. the SM prediction needs to be known to a comparable level to allow robust discrimination between the SM and BSM predictions.

In the right plot, we show $\delta_{\rm 2HDM}^{\rm align}$ in the fully degenerate case, that is $m_\phi\equiv \mH=\mA=\mHp$, in the $m_\phi$ vs.~$\laphi$ plane, with $\tan\beta=2$ and $\mbar$ fixed by the value of $\laphi$ (see~\cref{eq:thc}). Again, when the triple Higgs coupling $\laphi\sim0$, the 2HDM prediction corresponds to the SM, and a deviation from the SM is only possible when $\laphi\neq0$. Additionally, for the same value of the common triple Higgs coupling $\laphi$, the BSM effects decrease as the common Higgs mass $m_\phi$ increases. Overall, the deviations in the cross section are smaller than in the non-degenerate case, since it is more difficult to realise large triple Higgs couplings without making the theory non-unitary. In particular, in this degenerate scenario only $m_\phi \lesssim 750\ \mathrm{GeV}$ would lead to a deviation in the cross section potentially observable at the FCC-ee.

The results in this section show that the BSM effects induced in the Higgsstrahlung cross section by additional heavy Higgs bosons can be comparable in size to, or larger than, those expected from anomalous SM-like Higgs boson couplings to $Z$ bosons or from its self-coupling. Performing the computation in a full model rather than in an EFT circumvents the need of a matching, which proves to be particularly useful in situations where operators with mass dimension greater than six can give rise to sizeable contributions, see Refs.~\cite{Dawson:2022cmu,Ellis:2023zim,Adhikary:2025gdh} for related discussions.

Therefore, the interpretation of measurements from a high-precision FCC-ee $hZ$ run in terms of concrete UV-complete, yet phenomenologically flexible models (see also~\cite{Ramsey-Musolf:2021ldh}) will remain relevant going forward.

%%%%%%%%%%%%%%%%%%%%%%%%%%%%%%%%%%%%%%%
\subsection{Comments on Additional Information from Higgs Decays}
%%%%%%%%%%%%%%%%%%%%%%%%%%%%%%%%%%%%%%%
Different particle masses define distinct energy scales relevant to the scattering process, thereby increasing the number of independent measurements. This creates additional sources of sensitivity beyond variations in the centre-of-mass collision energy (see for instance~\cite{Bellafronte:2025jbk,Bellafronte:2026mhp}). To assess this additional sensitivity in our analysis, we include Higgs decays in Higgsstrahlung production to further disentangle the effects of Higgs mixing and inter-Higgs couplings. We consider only the dominant Higgs decay channel, $h \to b\bar{b}$. The one-loop corrections to the partial width $h \to b\bar{b}$ are computed, including electroweak and QCD corrections\footnote{The QCD corrections have been adapted from \texttt{HDECAY}~\cite{Djouadi:2018xqq}.}, using {\texttt{2HDECAY}}~\cite{Krause:2018wmo}. The electroweak corrections are obtained using the same renormalisation scheme as for $hZ$ production; the total width includes corrections only to two-body decays. As the branching ratio into bottom final states is the dominant one, corrections to the comparably smaller three- and four-body decays (namely the SM-like Higgs off-shell decays into massive gauge bosons) are relatively suppressed. Our approach should be considered a proxy for a full calculation. A complete treatment should also include off-shell effects~\cite{Altenkamp:2017kxk,Denner:2019fcr}, which are beyond the scope of this work. The resulting next-to-leading-order branching ratio estimate is combined with the $hZ$ production cross section using the narrow width approximation\footnote{The narrow-width approximation is absolutely justified here as the total width of the SM-like Higgs boson is of the order of $10^{-3}$~GeV.} as 
\begin{equation}
		\sigma(e^{+} e^{-} \to Zh \to Z b\bar{b})= \sigma(e^{+} e^{-} \to Zh )\times\mathrm{BR}(h \to b\bar{b})\,.
\end{equation}

We have cross-checked that, in the decoupling limit\footnote{The SM-like $\mathrm{BR}(h \to b\bar{b})$ is obtained in the decoupling limit with $m_{H}=m_{A}=m_{H^{\pm}}=\mbar=1.8\,\text{TeV}$ and $\cbma=0$. At NLO, we obtain $\mathrm{BR}(h \to b\bar{b}) =58\%$, which we use as the reference value in the plots.}, the LO and NLO branching ratios agree with the corresponding SM values for $h \to b\bar{b}$.

%%%%%%%%%%%%%%%%%%%
\begin{figure}
    \centering
    \includegraphics[width=0.99\linewidth]{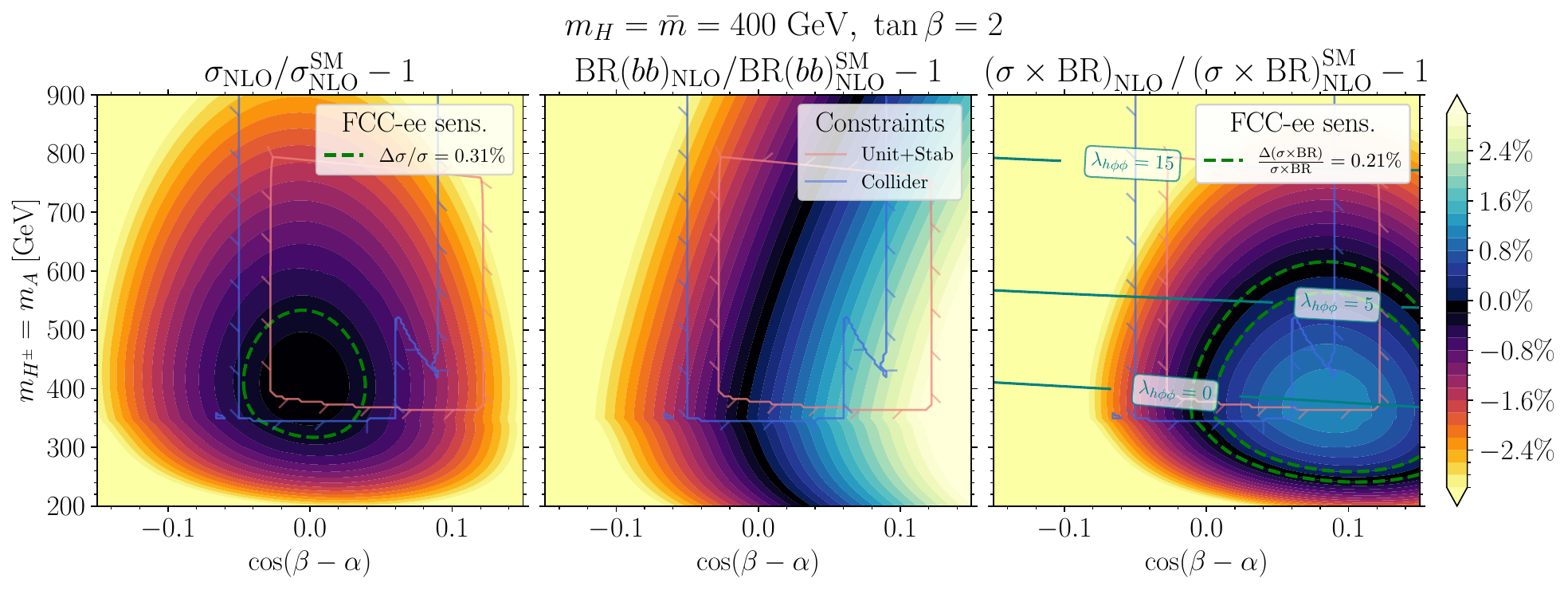}
    \includegraphics[width=0.99\linewidth]{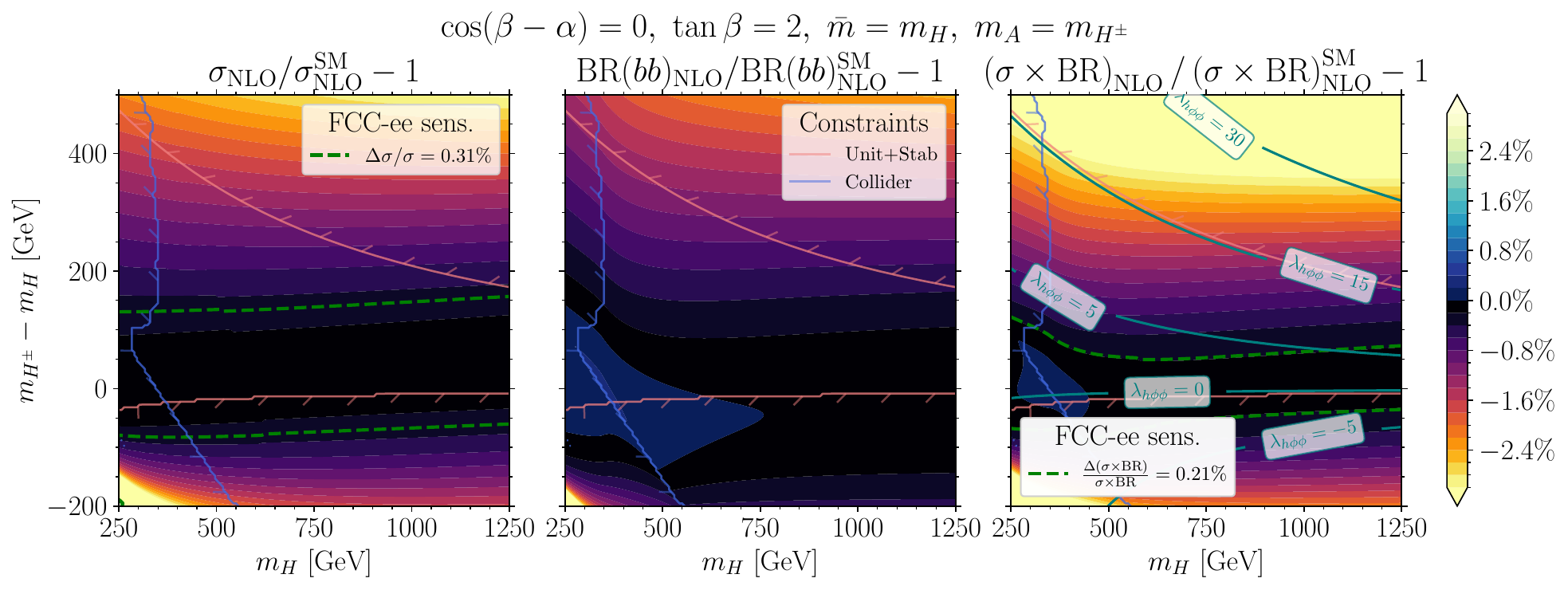}
    \caption{NLO deviations in Higgsstrahlung production and $\mathrm{BR}(h \to b\bar{b})$, as indicated by the color bars. The upper row corresponds to the benchmark scenario, $\mbar=m_{H}=400\ \mathrm{GeV}$ and $m_{A}=m_{H^{\pm}}$, while the lower row shows the deviations in the exact-alignment limit, $\cbma=0 $ with $\mbar=m_{H}$ and $m_{A}=m_{H^{\pm}}$. \textit{Left:} $\sigma(e^{+} e^{-} \to Zh)$ at $\sqrt{s}=240\ \mathrm{GeV}$. \textit{Middle:}  $\mathrm{BR}(h \to b\bar{b})$. \textit{Right:} Combined $\sigma(e^{+} e^{-} \to Zh \to Z b\bar{b})$. Parameter regions allowed by perturbative unitarity and vacuum stability are shown in pink, while those allowed by collider constraints are shown in blue. }
    \label{fig:XS-BR-planes}
\end{figure}
%%%%%%%%%%%%%%%%%%%

The effects are shown in~\cref{fig:XS-BR-planes} for two benchmark choices considered in the previous subsections. For comparison with the production result, the left panel shows the NLO deviations of the $hZ$ cross section relative to the SM. The middle panel shows the corresponding deviations in the NLO branching fraction, and the right panel shows the combined $\sigma(e^{+} e^{-} \to Z b\bar{b})$, together with the FCC-ee projection of $\pm0.21$\%~\cite{selvaggi_2025_n2emg-43f06}, shown by the dashed green lines. In the upper row, the effects of Higgs mixing and mass splitting $(m_\phi-\mbar)$ on NLO deviations in $h \to b\bar{b}$ are similar in order of magnitude to those observed in $hZ$ production. However, in contrast to production, the branching fraction shows both positive and negative deviations as $\cbma$ varies, i.e. it is enhanced for $\cbma>0$ and suppressed for $\cbma<0$. This can be traced back to the change in the value of the tree-level coupling of the SM-like Higgs boson to bottom quarks, see \cref{eq:zeta-f}, which in this case only depends on $\cbma$ (because $\tb$ is fixed) and changes sign at $\cbma=0$. It should also be noted that the effect of a non-SM value of $\kala$ is subdominant in the $h\to b\bar b$ decay~\cite{Degrassi:2016wml}. Therefore, the non-verticality of the contours in the cross section must arise from a loop-level effect related to NLO contributions from additional BSM bosons, which highlights their importance in the BR prediction. In addition, compared with production alone, including the decay slightly shifts the deviations in the full process $e^{+} e^{-} \to Z b\bar{b}$ towards positive values of $\cbma$.

\begin{figure}[t!]
    \centering
    \parbox{0.2\textwidth}{\includegraphics[width=1\linewidth]{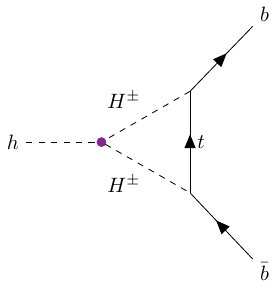}}
    \parbox{0.2\textwidth}{\includegraphics[width=1\linewidth]{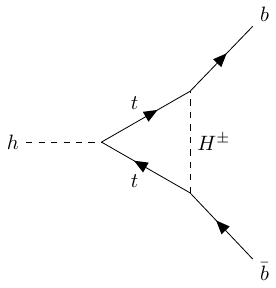}} \hfill
    \parbox{0.55\textwidth}{
    \caption{Main Feynman diagrams with additional non-SM-like scalars contributing at NLO to the Higgs decay to bottom-quark pairs in the alignment limit.
    Other BSM contributions may not vanish in the alignment limit, but they are further suppressed by the bottom mass.}
    \label{fig:FD-decay}}
\end{figure}

To isolate the effects of the BSM scalars, the second row of~\cref{fig:XS-BR-planes} shows the parameter space in the exact-alignment limit, $\cbma=0$. The constraints are shown in the $m_{H}$ vs. $m_{H^{\pm}}-m_{H}$ plane, considering $\mbar=m_{H}$ and $m_{A}=m_{H^{\pm}}$. We show in \cref{fig:FD-decay} a representative set of the main NLO diagrams that contribute to the BR and do not vanish in the alignment limit.
Notice that they are mediated by the charged Higgs boson $H^\pm$, as its Yukawa coupling introduces a term proportional to the top-quark mass.
Other diagrams would be comparably smaller because they would have at least one Yukawa coupling proportional to the bottom-quark mass. For positive mass splittings, $m_{H^{\pm}}-m_{H}>0$, the deviations in the branching fraction shown by the different contours in the middle panel become mostly flat at large $m_{H}$, particularly for $m_{H}>400\ \textrm{GeV}$, as it can be inferred from \cref{fig:XS-BR-planes}. The deviations are predominantly negative for large mass splittings. For example, for $m_{H^{\pm}}-m_{H}=400\ \textrm{GeV}$, the deviations can reach $-1.4\%$ at $m_{H} =400\ \textrm{GeV}$. In contrast, positive deviations occur for small mass splittings; for example, a deviation of $0.4\%$ is obtained for $m_{H^{\pm}}-m_{H}=60\ \textrm{GeV}$ at $m_{H}= 300\ \textrm{GeV}$. For negative mass splittings, the entire $m_{H}$ range is excluded by unitarity and stability constraints when $m_{H}-m_{H^{\pm}}>20\ \textrm{GeV}$. The $m_{H}$ dependence in the branching fraction for $m_{H}\simeq 400\ \textrm{GeV}$ is also visible in the full process shown in the right panel, whereas the deviations in the production cross section shown in the left plot are independent of $m_H$ over the entire range. The dependence on the mass splittings in the full process, however, remains qualitatively unchanged; large splittings lead to large negative deviations that exceed the expected FCC-ee precision of $\pm0.21\%$, as shown by the dashed green line. A positive deviation of approximately $0.2\%$, indicated by the dark blue contour, arises with smaller mass splittings and lower $m_{H}$ values. 
The observed dependence on the mass splittings can also be understood in terms of inter-Higgs couplings, $\lambda_{h\phi\phi}$, particularly the $hAA/hH^{\pm}H^{\mp}$ interactions. As discussed in the previous subsection (see~\cref{fig:heavyHiggs}), non-zero values of $\lambda_{h\phi\phi}$ also contribute to the NLO deviations. This follows from the explicit dependence of $\lambda_{h\phi\phi}$ on the mass splittings and $\mbar$ in the alignment limit, see~\cref{eq:thc}. From the plots in the second row, it can be inferred that varying mass splittings also modifies the corresponding trilinear couplings, with larger splittings leading to larger values of $\lambda_{h\phi\phi}$.

%%%%%%%%%%%%%%%%%%%
%%%%%%%%%%%%%%%%%%%

Overall, the effects of $\mathrm{BR}(h \to b\bar{b})$ modify the deviations in the full process relative to production alone. These deviations can be enhanced or reduced depending on whether the production and $b\bar{b}$ branching fraction contributions combine constructively or destructively. In the exact alignment limit, where the deviations arise solely from BSM scalars, including the Higgs decay makes the full process sensitive to smaller mass splittings. These effects also allow the FCC-ee sensitivity to be reached with smaller values of inter-Higgs BSM trilinear couplings.

Additional sensitivity to the effects of extra Higgs bosons in Higgsstrahlung may, in principle, be obtained from observables beyond the inclusive $Z$-associated Higgs-production cross section.  One possibility is the dependence on the centre-of-mass energy, motivated by the corresponding variation of the loop-induced coefficient $0.014\delta_\lambda$ in \cref{eq:kappa}. In the alignment limit, however, the contributions from the diagrams shown in \cref{fig:FD-scalars} generally exhibit only a mild dependence on the centre-of-mass energy. Measurements at different collision energies would therefore provide additional sensitivity primarily to the Higgs self-coupling, rather than to the effects of the extra scalars. A second possibility is to study differential cross-section distributions as a function of the scattering angle. In both cases, however, the observables do not appear to offer a strong means of discriminating the effects induced by the additional Higgs bosons. Whilst the corrections have a small non-trivial angular dependence, they are predominantly captured by the inclusive rate modification.

%%%%%%%%%%%%%%%%%%%%%%%%%%%%%%%%%%%%%%%
\section{Conclusions}
\label{sec:conc}
%%%%%%%%%%%%%%%%%%%%%%%%%%%%%%%%%%%%%%%
The current strategy for future colliders envisages a high-precision $e^+e^-$ collider, FCC-ee, with an extensive programme at the $Z$ pole and in associated $hZ$ production. In contrast to the LHC environment, FCC-ee will provide a comparatively small set of observables measured with exceptional experimental precision. This creates unprecedented opportunities for the indirect exploration of physics beyond the Standard Model.

Beyond generic EFT interpretations, this precision can be exploited within concrete renormalisable scenarios, allowing the measurements to be mapped consistently and efficiently onto a small number of parameters directly relevant to Higgs interactions. We focused on three main aspects: (i) the presence of additional, potentially heavy states relevant at the electroweak scale, (ii) modifications of its self-interactions, and (iii) the alignment of the 125 GeV scalar with the Standard Model Higgs direction. Using the 2HDM as a representative scenario incorporating all three effects, we have critically assessed the ability of FCC-ee to observe or constrain deviations from the Standard Model.

We find that, in the presence of Higgs mixing, additional states remain relevant as modifiers of the $hZ$ production cross section, independently of changes to the Higgs self-coupling. In particular, the exotic scalar contributions remain significant in alignment without decoupling. Consequently, a direct interpretation of the $hZ$ cross section solely in terms of $\kappa_\lambda$ is generally not meaningful when the aim is to draw conclusions about realistic extensions of the Standard Model. Nevertheless, FCC-ee retains substantial combined sensitivity to extended scalar sectors. Its sensitivity can therefore place important constraints on the structure of TeV-scale physics, despite the comparatively small number of independent observables available relative to hadron colliders.

%%%%%%%%%%%%%%%%%%%
\subsection*{Acknowledgements}
%%%%%%%%%%%%%%%%%%%
A., S.D.N., and M.M. acknowledge support by the Deutsche Forschungsgemeinschaft (DFG, German Research Foundation) under grant 396021762 --- TRR 257.
F.A.~acknowledges support by the Deutsche Forschungsgemeinschaft (DFG, German Research Foundation) under Germany's Excellence Strategy -- EXC 2121 ``Quantum Universe'' -- 390833306. The work of F.A.~has also been partially funded by the Deutsche Forschungsgemeinschaft (DFG, German Research Foundation) -- 491245950.
%%%%%%%%%%%%%%%%%%%
\bibliographystyle{JHEP}
\bibliography{references}
%%%%%%%%%%%%%%%%%%%
\end{document}